\documentclass[apl,preprint,reprint,superscriptaddress]{revtex4-1}
\usepackage{graphicx}
\usepackage{dcolumn}
\usepackage{bm}

\begin{document}
\title{Production yield of rare-earth ions implanted into an optical crystal}




\author{Thomas Kornher}
\email{t.kornher@physik.uni-stuttgart.de}
\author{Kangwei Xia}

\author{Roman Kolesov}
\affiliation{3rd Physics Institute, Stuttgart University and 
Stuttgart Research Center of Photonic Engineering (SCoPE)}
\author{Nadezhda Kukharchyk}
\affiliation{Angewandte Festk\"{o}rperphysik, Ruhr-Universit\"{a}t Bochum, 
D-44780 Bochum, Germany}
\author{Rolf Reuter}
\affiliation{3rd Physics Institute, Stuttgart 
University and 
Stuttgart Research Center of Photonic Engineering (SCoPE)}
\author{Petr Siyushev}
\affiliation{Universit\"{a}t Ulm,
Institut f\"{u}r Quantenoptik, D-89081 Ulm,
Germany}
\author{Rainer St\"{o}hr}
\affiliation{3rd Physics Institute, Stuttgart University and 
Stuttgart Research Center of Photonic Engineering (SCoPE)}
\affiliation{Institute for Quantum Computing, University of Waterloo, 
Waterloo,
Ontario, Canada}
\author{Matthias Schreck}
\affiliation{Experimentalphysik IV,
Universit\"{a}t Augsburg,
D-86159 Augsburg,
Germany}
\author{Hans-Werner Becker}
\affiliation{RUBION, Ruhr-Universit\"{a}t Bochum, D-44780 Bochum, Germany}
\author{Bruno Villa}
\affiliation{3rd Physics Institute, Stuttgart University and 
Stuttgart Research Center of Photonic Engineering (SCoPE)}
\author{Andreas D. Wieck}
\affiliation{Angewandte Festk\"{o}rperphysik, Ruhr-Universit\"{a}t Bochum, 
D-44780 Bochum, Germany}
\author{J\"{o}rg Wrachtrup}
\affiliation{3rd Physics Institute, Stuttgart University and 
Stuttgart Research Center of Photonic Engineering (SCoPE)}

\begin{abstract}
Rare-earth ions doped into desired locations of optical crystals might 
enable a range of novel integrated photonic devices for quantum applications. 
With this aim, we have investigated the production yield of cerium and 
praseodymium by means of ion implantation. As a measure, the collected 
fluorescence intensity from both, implanted samples and single centers was used. 
With a tailored annealing procedure for cerium, a yield up to 53\,\% was 
estimated. Praseodymium yield amounts up to 91\,\%. Such high implantation yield 
indicates a feasibility of creation of nanopatterned rare-earth (RE) doping and 
suggests strong potential of RE species for on-chip photonic devices.
\end{abstract}

\maketitle
Crystals doped with RE ions, well known due to their key 
application in laser technology, additionally aspire to become a viable 
contender
in solid-state quantum 
information processing. With landmark achievements like coherent manipulation 
\cite{siyushev2014coherent} and
all-optical addressing of a single ion spin \cite{xia2015all}, up to six 
hours long storage times of quantum states \cite{zhong2015optically}, 
and quantum memory for entangled photon pairs \cite{clausen2011quantum}, RE 
doped crystals show strong potential in quantum optics research and technology.

Typically, crystals are doped during their growth 
to generate optically detectable RE ensembles \cite{bottger2009effects}. 
Scalable, integrated quantum networks, however, require nanoscopic engineering 
of RE ions, in order to employ them as stationary qubits. Ion implantation as 
a means for controlled doping of crystals is therefore a prerequisite for a 
more 
versatile implementation in experiments, as was the case for nitrogen vacancy 
centers in diamond \cite{li2015coherent, fairchild2008fabrication, 
neumann2010quantum, dolde2013room, riedrich2015nanoimplantation}.

In this Letter, we report on the creation efficiency of trivalent cerium and 
praseodymium ions in yttrium aluminum garnet (YAG) doped by ion implantation. A 
wide range of 
ion fluences and implantation energies was used in order to 
obtain a comprehensive picture of the production yield of two promising  
RE ion species in YAG. Also, post-implantation 
annealing atmospheres were investigated and an advantageous approach for yield 
estimation is presented. The motivation behind these experiments is to optimize 
the 
generation of fluorescent 
RE ions in crystals by means of ion implantation
\cite{pezzagna2010creation, IMPEr3YSO, IMPNVcenters2}.
\\

Trivalent RE ions can easily substitute yttrium ions 
in 
the crystal lattice of YAG, thus forming color centers featuring optical 
transitions with high quality factor. Foundation of these high-Q transitions 
are electrons located in 
the partially filled 4f shell of RE ions, which are shielded from the 
environment by closed 
outer 5s and 5p shells. This results in long coherence times of both electron 
spin and nuclear spin. In the 
experiment, 
cerium and praseodymium ions are used as dopants, and their fluorescence 
intensity 
is detected in a home-built 
high resolution confocal and upconverting microscope setup, respectively. 

\begin{figure}
\includegraphics[width=0.48\textwidth]{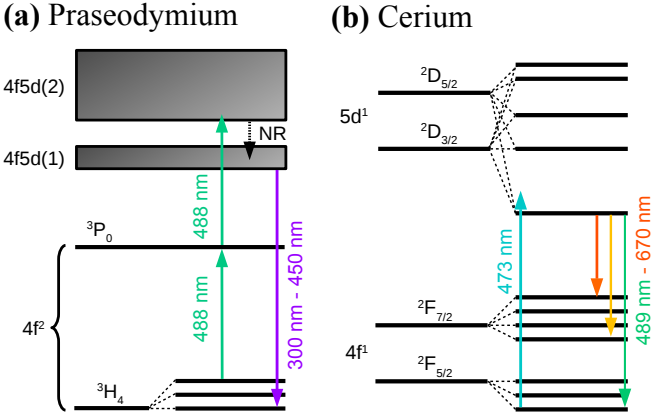}
\caption{Electronic level structures of (a) praseodymium and (b) cerium with 
employed excitation and emission wavelengths. NR: 
non-radiative decay.\label{fig:level}}
\end{figure}

Trivalent praseodymium 
ions in YAG are excited by a two-step upconversion process \cite{ganem1992one} 
with a 
diode laser of 488.25\,nm wavelength. The first excitation step involves a 
spectrally 
narrow, 
parity forbidden 4f-4f transition from $^3$H$_4$ ground state to 
$^3$P$_0$ state, as depicted in 
fig.\,\ref{fig:level}\,(a) \cite{gruber1989symmetry}. In $^3$P$_0$, the electron 
exhibits 
a 
lifetime of 8\,$\mu$s, during which it is able to absorb another photon and 
thus is promoted into the 4f5d(2) band, where non-radiative decay onto the 
lowest 
4f5d(1) level occurs. The 4f5d shell enables parity-allowed optical transitions 
with a lifetime of approximately 18\,ns to 4f states, featuring ultraviolet 
fluorescence detected in a spectral range of 
290-370\,nm \cite{gayen1983two}. Due to a high cycling 
rate, this scheme previously allowed detection of single Pr$^{3+}$ 
centers in YAG \cite{kolesov2012optical}. 

As shown in fig.\,\ref{fig:level}\,(b), trivalent cerium ions 
in YAG are non-resonantly 
excited with a diode laser of 473\,nm wavelength, thus pumping the 4f$^1$ 
ground level to 
the 
lowest 5d$^1$ level, which exhibits a lifetime of 60\,ns \cite{Ce3YAGESA}. 
Its strong phonon-sideband emission is related to the 5d-4f transition, 
which is detected in a 
491-630\,nm spectral window. Quantum 
efficiencies of these transitions are close to unity 
\cite{weber1973nonradiative}. 
Single ion detection of both cerium and praseodymium plays a key role in 
fluorescence yield 
estimation \cite{kolesov2013mapping, kolesov2012optical}. 

\begin{figure}
\includegraphics[width=0.48\textwidth]{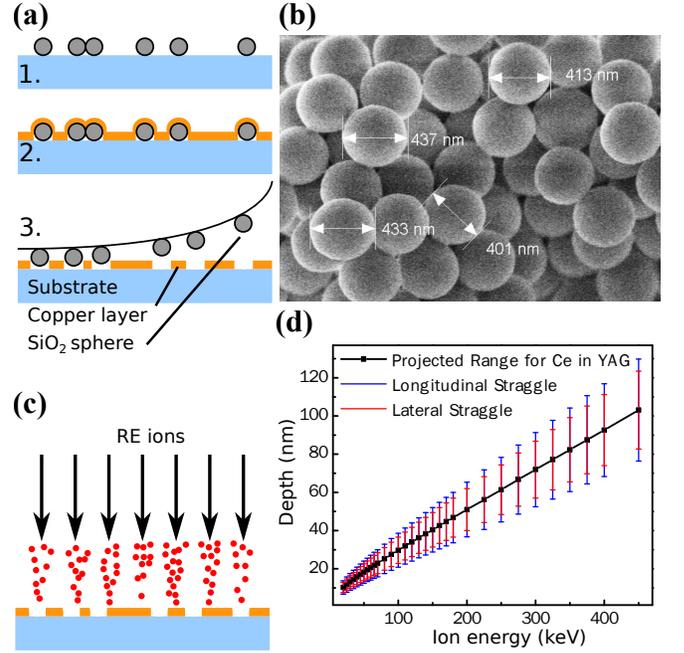}
\caption{(a) Mask-making process is illustrated in three steps.  (b) SEM image 
of stacked SiO$_2$ nano-spheres to measure their size. (c) Area implantation 
through mask. (d) Simulated 
implantation depths of cerium implanted into YAG with corresponding 
straggles. \label{fig:IMPL11}}
\end{figure}

To obtain a robust yield estimation, following sample processing was carried 
out. Before implantation, samples are covered with a perforated copper mask.
For that, monodisperse SiO$_2$ nanospheres (SEM image shown in figure 
\ref{fig:IMPL11} (b), radius 
$r_{\text{sphere}}\approx210$\,nm) 
are spin-coated onto the polished surface of the crystal. The nanospheres were 
prepared by a sol-gel method as described in the Supplementary Material. A 
subsequent copper evaporation step results in a 200\,nm thick copper layer. 
Then, the SiO$_2$ spheres are removed, leaving 
uniform holes in the copper mask, as illustrated in figure \ref{fig:IMPL11} 
(a). RE ions were subsequently implanted through the mask, depicted in figure 
\ref{fig:IMPL11} (c). After implantation, the copper mask is removed from the 
crystal by wet etching 
in FeCl$_3$ solution. Samples are then 
annealed in different atmospheres at 1200$^{\circ}$C for 24 
hours to heal out implantation induced damage. 
Praseodymium-implanted samples are annealed in air. For 
cerium, previously conducted studies suggest a reducing 
atmosphere \cite{ACTCeYAGARH2, ACTCeYAGN2H2nano, ACTCeYAGCO2} to 
improve stabilization in the desired charge state in the 
crystal. Our preliminary experiments confirmed this behavior in a reducing 
atmosphere of argon and hydrogen 
(95\,\%/5\,\%) when compared to an inert argon atmosphere. In previous work 
\cite{xia2015all}, we reported gradual bleaching of 
Ce$^{3+}$ centers under continuous wave (CW) excitation, while they are 
photostable under femtosecond illumination. On the contrary, cerium ions in the 
samples annealed under Ar$\,+\,\text{H}_2$ atmosphere are photostable under CW 
excitation and, therefore, allowed us to use CW diode laser for the optical 
studies.

We used an EIKO E-100 focused 
ion beam (FIB) system for implantation, where ions are extracted from a 
home-made liquid metal ion source (LMIS) \cite{IMPLMISref2}, containing an 
alloy of either cerium or praseodymium. The 
implantation energy ranged 
between 75\,keV and 300\,keV and determined the expected depths of 
implanted ions. Figure \ref{fig:IMPL11}\,(d) shows SRIM 
simulations \cite{ziegler2010srim} 
regarding ion 
depths, together with longitudinal straggle and lateral straggle depending 
on the 
implantation energy. Going to lower implantation energies can dramatically 
decrease the straggle volume and results in more precise nanoscale 
engineering. 

\begin{figure}
\vspace{0.1cm}
\includegraphics[width=0.48\textwidth]{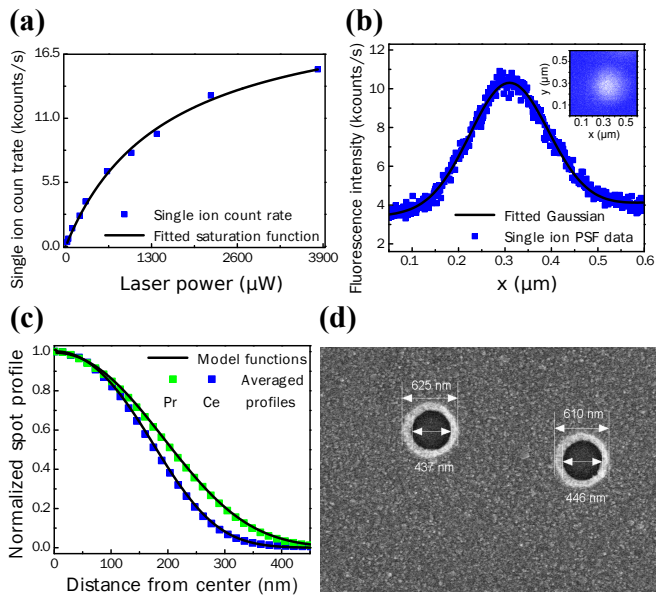}
\caption{(a) Cerium power study on a single center. The corresponding fit yields 
a saturation laser power of I$_0 = 1500$\,mW. Accordingly, fluorescent 
measurements were taken at 300\,mW, well below this threshold to avoid 
saturation
effects. (b) Measured PSF of a 
single cerium ion and corresponding Gaussian fit. Inset: 2-D laser scanning 
microscope image  of a single fluorescent cerium ion. (c) Averaged spot 
profiles with 
corresponding spot model fits. (d) SEM image of the copper 
mask. \label{fig:Yield1}}
\end{figure}

Ion fluence ranged between $
10^{12}$\,\,ions/cm$^{2}$ and $
10^{14}$\,\,ions/cm$^{2}$. This results in a dynamic range of up to five 
orders 
of magnitude, as single ion fluorescence needs to be compared to up 
to $\sim$\,130000 fluorescing ions in one spot. In order to overcome 
limitations imposed 
by the fluorescence detector and to avoid center saturation, the laser needs to 
be operated in the linear excitation power range of the respective ion species 
and the linear intensity range for the detector. 
Therefore, power studies were carried out. For cerium, a two-level 
rate equation model was used to fit measured data, as shown in 
figure \ref{fig:Yield1}\,(a). Solution to the two-level approach is the laser 
power dependent single center fluorescence intensity 

\begin{equation}
F_{\text{sc}}(I) = A \cdot \frac{I}{I + I_0},
\label{eq:rateEQ}
\end{equation}
where $A$ is maximum fluorescence intensity of a single ion and $I_0$ is the
saturating laser power.

Due to the two-step upconversion in the 
case of praseodymium, ion 
saturation is far less likely, so that only detector saturation was monitored 
in the power study, shown in the Supplementary Material.


For production yield estimation of implanted RE ions, we modeled the 
emission of implanted spots and compared it to the measured 
emission. Modeling of the spot emission is based on the single 
RE ion point spread function (PSF). The inset in figure \ref{fig:Yield1}\,(b) 
displays the 
corresponding 2-D laser scanning microscope image of a single cerium ion.  
By fitting a Gaussian function to a cross 
section scan of a single RE ion, we obtain the PSF radius, shown in 
figure \ref{fig:Yield1}\,(b). 
The emission profile of the implanted spot is given by the convolution of the 
PSF of the microscope with the distribution of the implanted ions:
\begin{equation}
F_0\left(\mathbf{r}^{\prime}\right)=\int \mathrm{d} \mathbf{r} \cdot \mathbf{r} 
\varrho 
\left(\mathbf{r}\right) I_{\text{PSF}}\left( \mathbf{r}-\mathbf{r^{\prime}} 
\right)
\end{equation}
where $\varrho(\mathbf{r})$ is the spatial distribution of the implanted ions 
and $I_{\text{PSF}}(\mathbf{r})$ the characteristic emission of a single ion. 
More detailed description of modeling $\varrho(\mathbf{r})$ is given in the 
Supplementary Material. Spot emission profiles were 
extracted from fluorescent scans of implanted spots, 
averaged and fitted with $F_0(\mathbf{r}')$ as depicted in figure 
\ref{fig:Yield1}\,(c). Furthermore, a copper mask characterization was done 
with secondary electron 
microscopy (SEM), where holes were found to also feature a rim, as depicted in 
figure \ref{fig:Yield1}\,(d). However, implanted RE ions can only penetrate 
through the inner part of the rim, where the copper layer is thinner than 
the energy-dependent penetration depth of RE ions into copper. For energies of 
$75-300$\,keV, the penetration depth ranges between $15-45$\,nm according to 
SRIM simulations. 
Consequently, rim widths obtained directly through SEM measurements decrease 
to the effective rim width $r_{\text{rim}}$. Holes were found to have a radius 
$r_{\text{hole, SEM}} = 221\pm3$\,nm, with a rim of an effective 
width $r_{\text{rim, SEM}} = 15\pm5$\,nm. For a comparison, spot profiles 
obtained from optical measurements of both, praseodymium-implanted and 
cerium-implanted samples were fitted with the introduced model. As a result, 
the modeled hole radius amounts to $r_{\text{hole, fit}} = 206\pm19$\,nm and 
the modeld rim width to $r_{\text{rim, fit}} = 5\pm5$\,nm. Our approach can
confirm the copper mask parameters measured by SEM.

\begin{figure}
\vspace{0.5cm}
\includegraphics[width=0.48\textwidth]{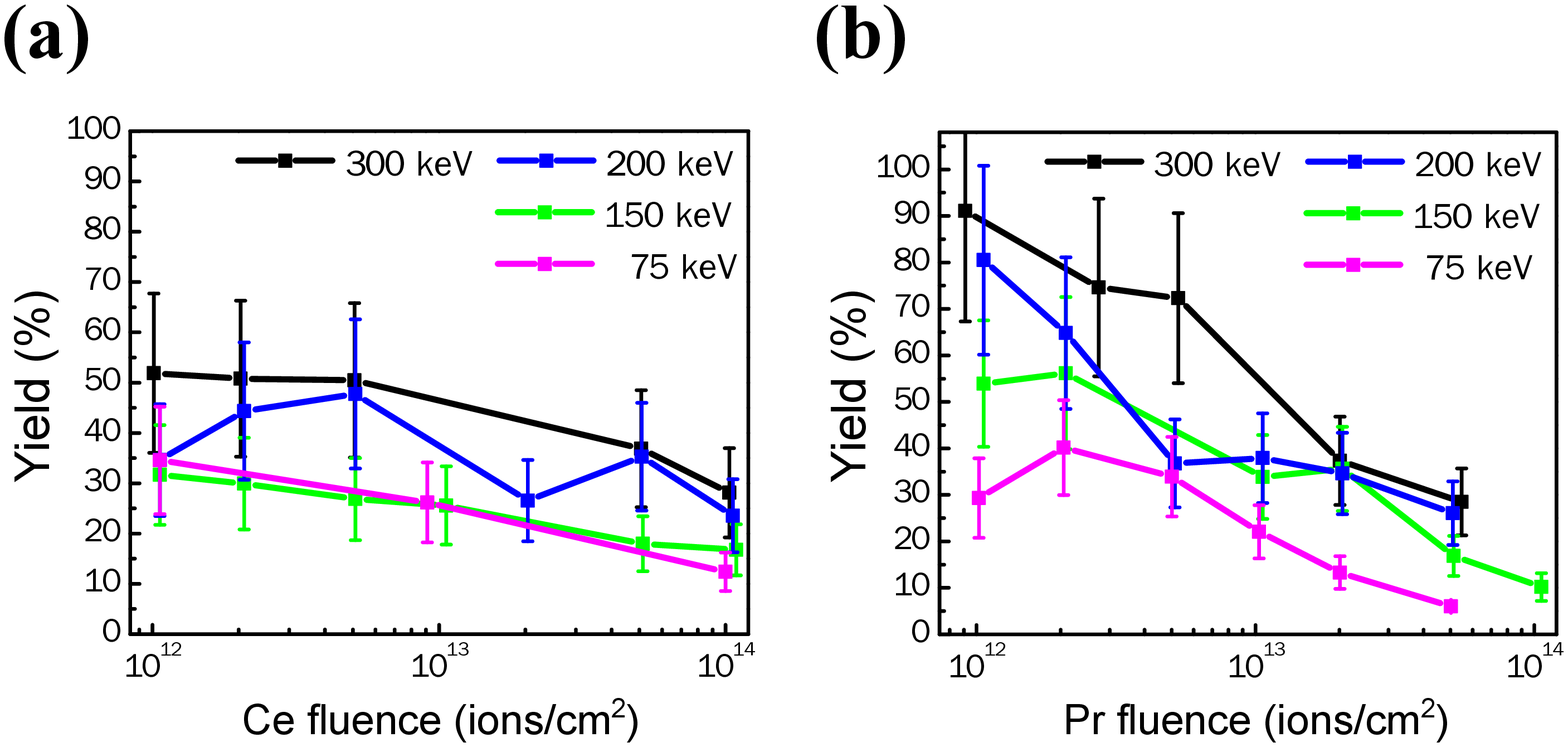}
\caption{Production yield of implanted (a) cerium ions and (b) praseodymium 
ions for four 
different implantation energies. \label{fig:Yield2}}
\end{figure}

Figures \ref{fig:Yield2} (a) and (b) show the efficiency of the produced RE 
ions in their 
fluorescing RE$^{3+}$ charge state for different implantation energies 
as a function of the ion fluence. The figures show a decrease in yield 
with increasing fluence of implanted ions. Higher implantation energies feature 
a 
higher yield. For higher energies, ions exhibit a larger straggle, longer 
travel distance in the crystal and they also generate more defects per 
implanted ion. All 
these effects can contribute to a higher possibility for implanted RE ions to 
settle in the crystal lattice and to be activated after annealing.
The fluence-dependent behavior may be explained by an increase in local RE ion 
density with 
increasing ion fluence. As the straggle volume for each ion 
energy stays constant for a varying fluence, the final amount of RE ions within 
 the same volume is fluence-dependent. The more 
ions are implanted in such a space, the less likely it is for the individual 
ion 
to find a proper location to replace yttrium. Peak production yield values are 
91\,\% for praseodymium and 53\,\% for cerium for the lowest ion fluence 
in each case. In principle, both values can reach unity, provided ideal 
activation procedures are found.

In conclusion, this quantitative study confirms high production yield values 
for implanted RE ions. Reported values of other color centers, 
such as silicon vacancy centers in diamond show a yield of 15\,\% with an 
implantation energy of 60\,keV \cite{tamura2014array}. Investigation into 
nitrogen vacancy center generation in diamond meanwhile reached production 
yield values of 25\,\% for implantation energies between 2.5\,keV and 
20\,keV \cite{antonov2014statistical} and almost 50\,\% for 
implantations at MeV energies \cite{pezzagna2010creation}. Unreached is the 
activation rate of praseodymium, with almost unity yield. This suggests single 
ion implantation \cite{schnitzler2010focusing} 
attempts to become feasible. In turn, high spectral stability of the optical 
lines of the implanted RE$^{3+}$ ions reported in our previous work 
\cite{xia2015all} makes them very favorable candidates for optically 
addressable single ion qubits. This work also paves the way toward low energy 
implantations in the range of 
$0.1-10$\,keV, which would result in deterministic high resolution 
nano-positioning of RE ions. Under which conditions a high production yield for 
RE ions can be maintained for such low energy implantations has yet to be 
investigated.
\\

The work was financially supported by ERC SQUTEC, EU-SIQS SFB TR21 and DFG 
KO4999/1-1. N.K. and A.D.W acknowledge gratefully support of Mercur  
Pr-2013-0001, DFG-TRR160,  BMBF - Q.com-H  16KIS0109, and the DFH/UFA  
CDFA-05-06.


%
%

%



\bibliography{Bibtexfile141118}

\end{document}